# Models simulation and interoperability using MDA and HLA

Hind El Haouzi

**CRAN (UMR 7039), University Henry Poincaré Nancy I**

*F 54506 Vandoeuvre les Nancy, France*
*Hind.elhaouzi@ cran.uhp-nancy.fr,*

ABSTRACT: *In the manufacturing context, there have been numerous efforts to use modeling and simulation tools and techniques to improve manufacturing efficiency over the last four decades. While an increasing number of manufacturing system decisions are being made based on the use of models, their use is still sporadic in many manufacturing environments. Our paper advocates for an approach combining MDA (model driven architecture) and HLA (High Level Architecture), the IEEE standard for modeling and simulation, in order to overcome the deficiencies of current simulation methods at the level of interoperability and reuse.*

RÉSUMÉ : *L'évolution des entreprises vers des architectures de plus en plus complexes a rendu incontournable les techniques de modélisation et de la simulation pour l'aide à la décision. Cependant, leur utilisation reste encore sporadique et manque de méthodologies pour construire des modèles de simulation réutilisables et garantir leur interopérabilité dans le contexte de la simulation distribuée. Nos travaux visent à combiner les architectures MDA et HLA pour répondre à ces challenges.*

MOTS-CLÉS: *interopérabilité, décisions distribuées vs centralisées, simulation à événements discret.*

**1. Introduction**

Today manufacturing systems need to be reactive to internal disturbances (e.g. machine breakdown) as well as external disturbances (e.g. economic changes, changing demand, and product adaptation). Consequently research in manufacturing system control has moved away from traditional centralized approaches where decision making is hierarchically broadcast from the higher decisional levels down to the operational units to more distributed architectures. In this way, hierarchical architectures promote production control by distributing every decision capacities in autonomous entities, without any centralized view of the shop floor status. To ensure consistency of decision making, more pragmatic approaches are based on hybrid control which combines the predictability of the centralized control with the agility and robustness against disturbances of the distributed control.

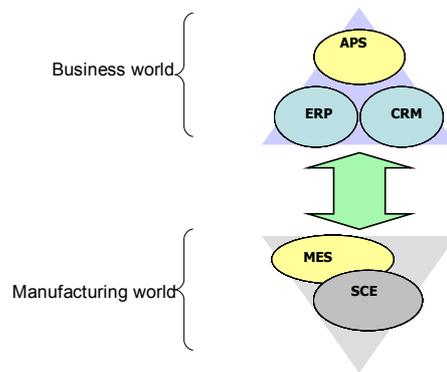

**Figure. 1**. *Business and manufacturing separation*

In this paper, we assume that the enterprise environment is composed of two separated worlds (see ): *(i)* On one hand, a business world in which centralized decisions concerning the whole enterprise are taken, *(ii)* On the other hand, a manufacturing world responsible of distributed real time decisions in order to control and execute manufacturing processes of shop-floor level. This separation is due to several reasons; the most important is the difference of objectives and rules used in the decision making process. In the figure 1, we tried to position some of the major applications in the enterprise: Manufacturing Execution System (MES), Supply Chain Execution (SCE), Advanced Planning Scheduling system (APS), Enterprise Resource Planning (ERP), Customer Relationship Management (CRM).

The separation of the two worlds implies separation of the models representing each world. Thus, we obtain models representing different universes of discourse (business/manufacturing), and using different concepts, different modeling rules and concerns.

In the manufacturing context, there have been numerous efforts to use modeling and simulation tools and techniques to improve manufacturing efficiency over the last four decades. While an increasing number of manufacturing system decisions are being made based on the use of models, their use is still sporadic in many manufacturing environments. We believe that there is a real need for pervasive use of modeling and simulation for decision support in current and future manufacturing systems. There are several challenges that need to be addressed by the simulation community to achieve this vision.

In this paper, we highlight two major challenges: models interoperability (synchronization, coordination and coherence) and reusability in the decision taking process. To handle these challenges, several issues are to be solved; for example, data integration, time management and synchronization between different simulation models (distributed or centralized).

## 2. Towards Simulation

In fact, models are intended to support management decisions about the system and a single model will often not be capable of supporting all decisions. Rather, different decisions require different models because various aspects of the design and operation of the system will be important for the questions being asked. While spreadsheet and queuing models are useful for answering basic questions about manufacturing systems, discrete event simulation models are often needed to answer more specific questions about how a complex manufacturing system will perform [1]. In general, simulation is a practical methodology for understanding the high-level dynamics of a complex manufacturing system. According to [2], simulation has several strengths including:

– Time compression: the potential to simulate years of real system operation in a much shorter time.

– Component integration: the ability to integrate complex system components to study their interactions,

– Risk avoidance: hypothetical or potentially dangerous systems can be studied without the financial or physical risks that may be involved in building and studying a real system.

– Physical scaling: the ability to study much larger or smaller versions of a system.

– Repeatability: the ability to study different systems in identical environments or the same system in different environments.

– Control: everything in a simulated environment can be precisely monitored and exactly controlled.

In our context, complex simulations involve individual simulations of several different types of systems (Business and Manufacturing), combined with other aspects of the total environment to be simulated (such as interactions…). Often simulations of some of these components already exist, having been developed for a different purpose, and they could be used in a new simulation. Unfortunately, it is often necessary to make extensive modifications to adapt the component simulation model so that it can be integrated into a new combined simulation. Thus, traditional simulation models often lack two desirable properties: reusability and interoperability.

Reusability, as the name suggests, means that component simulation models can be reused in different simulation scenarios and applications Interoperability implies an ability to combine component simulations on distributed computing platforms of different types, often with real-time operation.

## 3. The federation mechanism

### 3.1. *A theoretical approach*

In order to enable the simulation of all different models of the enterprise as a whole, we need to establish a global federated model which will represent the most important aspects and concepts of each specific partial model of the enterprise. This federated model is an instantiation of a federated meta-model where concepts from different universes of discourse can coexist (business, manufacturing, etc.).

Several models interoperability and federation approaches exist; UEML [3]. is one of these approaches. The main problem with the use of UEML in this context, is that in order to achieve our first objective which is simulation, we should be able to use UEML models in simulation tools, and this is not possible currently. To encompass this problem, in this section, we introduce a generic approach for interoperability based on a model driven architecture (MDA) [4]. Figure 2 shows the four-level ontological approach levels for modeling that are used in the MDA. As it is explained in [5], the lowest level $M^0$ presents different subjects for modeling, called universe of discourse. The level $M^1$ contains different models of each universe of discourse. The next level $M^2$ presents domain specific meta-models: one meta-model for each of the domains of interest relevant for the $M^1$ models. And finally, $M^3$ level presents a meta-meta-model designed to allow the definition of all the existing in the scope of the meta-models.

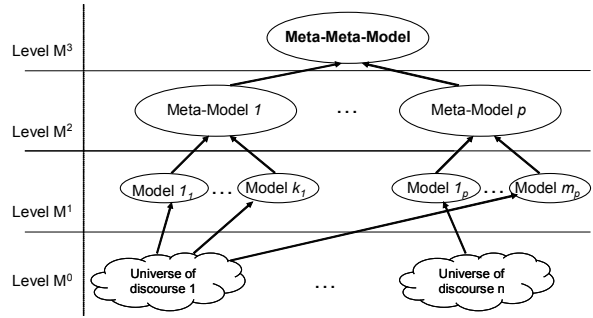

**Figure. 2** *The four-level ontological approach..*

In case of MDA, each application can be considered as a specific use of a model defined in the $M^1$ level which is based on meta-model defined in $M^2$. Application interoperability can then be resolved either by interconnecting applications together using a level $M^0$ exclusive reasoning, or by establishing a top-down approach for resolving applications interoperability based on the four levels of the MDA. Several research works have been done in order to resolve meta-models mapping problems. In [6], we explain how models transformation can be resolved by establishing transformation rules between meta-models. Transformation rules define a mapping that guides model transformations from the instances of the source meta-model to instances of the target meta-model.

To ensure interoperability between applications that handle those different meta-models, we should first define mappings that enable transforming one instance of a meta-model in an instance of another meta-model. Let us consider two applications A and B; A and B are interoperable, if and only if there is a mapping from the meta-model of A ($M_A$) to the meta-model of B ($M_B$) and a mapping form $M_B$ to $M_A$. Those mappings ensure that we can build a model compatible with A from a model used by B (and vice versa).

The federation mechanism in the MDA context consists on defining a unified meta-model in the M2 level. This federated meta-model is the interoperability gateway between concepts defined in different universes of discourse, in our case the business and manufacturing worlds.

### 3.2. High level Architecture:

In this section, we describe an implementation approach of simulation models. As said previously, within the new manufacturing system context, two major problems are encountered by simulation developers; reusability and interoperability. In simulation literature the High Level Architecture [7] seems the one partial solution to those challenges. In the HLA, each simulation or other software system is run as a separate federate (process), and the collection of all federates is called a federation. Each federate can be developed independently and implemented using different software languages and different hardware platforms.

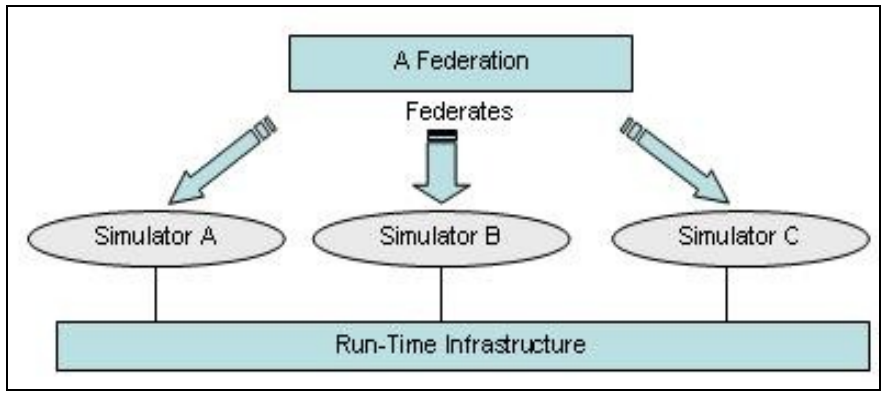

**Figure. 4** *The four-level ontological approach..*

The HLA Baseline Definition was completed in 1996 and was adopted as the Facility for Distributed Simulation Systems 1.0 by the Object Management Group (OMG) in 1998. The HLA was approved as an open standard through the Institute of Electrical and Electronic Engineers (IEEE) - IEEE Standard 1516

The HLA provides a general framework within which simulation developers can structure and describe their simulation applications. It consists of three components: Federation Rules, Interface Specification, and Object Model Template (OMT).

*3.2.1 Federation Rules*

At the highest level, the HLA consists of a set of 10 HLA rules which must be obeyed if a federate or federation is to be regarded as HLA compliant. The HLA rules are divided into two groups consisting of 5 rules for HLA federations and 5 rules for HLA federates. The federation rules establish the ground rules for creating a federation, including documentation requirements, object representation; data interchange interfacing requirements and attribute ownership. The federate rules deal with the individual federates. They cover documentation, control of and transfer of relevant object attributes, and time-management.

*3.2.2 Interface Specification*

The interface specification defines a standard for a Run-Time Infrastructure (RTI). The RTI is software that conforms to the specification but is not itself part of the specification. It provides the software services (Time management, Federation management…), which are necessary to support an HLA compliant simulation. Different versions of the RTI are possible but it difficult to find Open source and free version.

*3.2.3 The Object Template Model*

Reusability and interoperability require that all objects and interactions managed by a federate, and visible outside the federate, should be specified in detail and with a common format. The Object Model Template (OMT) provides a standard for documenting HLA Object Model information and defines the Federation Object Model (FOM), the Simulation (or Federate) Object Model (SOM) and the Management Object Model (MOM).

**4. On-going work**

This paper presents ongoing work of the first year PhD student Hind El Haouzi in collaboration with Phd Student Salah Baïna who works essentially on enterprise models interoperability using the MDA approach. The paper does not expose solutions but tries essentially to highlights major problems encountered in the domain of distributed models simulation.

On-going work aims to combining both approaches MDA and HLA to obtain a framework (Analysis and technical tools) for ensuring interoperability between Business and Manufacturing Models. The discrete-event simulation will be used to evaluate decision impacts in both levels. As a case study, the framework will be a useful support to integrate Identification technologies (RFID) in Trane group legacy system.

**5. References**

Chance, F., Robinson, J., and J. Fowler, "Supporting manufacturing with simulation: model design, development, and deployment", Proceedings of the 1996 Winter Simulation Conference, San Diego, CA, 1996, pp. 1-8.

Yücesan, E. And Fowler, J. ,"Simulation Analysis of Manufacturing and Logistics Systems", Encylclopedia of Production and Manufacturing Management, Kluwer Academic Publishers, Boston, P. Swamidass ed. , pp. 687-697., 2000

Berio G., /et al./ 2003 D3.2: /Core constructs, architecture and development strategy/, UEML TN IST – 2001 – 34229, March 2003.

Mellor S.J., Kendall S., Uhl A. and Weise D. *Model Driven Architecture,* Addison-Wesley Pub Co, March, ISBN: 0201788918.

Naumenko, A. Wegmann, A. « Two Approaches in System Modelling and Their Illustrations with MDA and RM-ODP ». In *ICEIS 2003, the 5th International Conference on Enterprise Information Systems* (2003), p. 398-402.

Lemesle, R. « Transformation Rules Based on Meta-Modelling ». *EDOC'98,* La Jolla, California, 3-5 November 1998, p. 113-122.

The Institute of Electrical and Electronics Engineers, Inc., IEEE Std 1516-2000. IEEE Standard for Modeling and Simulation (M&S) High Level Architecture (HLA) Framework and Rules. 11 December 2000.